\documentstyle[preprint,aps]{revtex}
\tightenlines
\begin{document}

\title{Constraints on $AdS_5$ Embeddings}

\author{\normalsize{Philip D. Mannheim\footnote{Email address: 
mannheim@uconnvm.uconn.edu}} \\
\normalsize{Center for Theoretical Physics, Laboratory for Nuclear
Science and Department of Physics,
Massachusetts Institute of Technology, Cambridge, Massachusetts 02139} \\
\normalsize{and} \\
\normalsize{Department of Physics,
University of Connecticut, Storrs, CT 06269\footnote{permanent address}} \\
\normalsize{(MIT-CTP-2989, hep-th/0009065 v2, March 5, 2001)} \\}

\maketitle

\begin{abstract}
We show that the embedding of either a static or a time dependent 
maximally 3-symmetric brane with non-zero spatial curvature $k$ into a
non-compactified $AdS_5$ bulk does not yield exponential suppression of 
the geometry away from the brane. Implications of this result for 
brane-localized gravity are discussed.
\end{abstract}

\section {Static Branes}

It was recently pointed out \cite{Randall1999a,Randall1999b} that if our 
4-dimensional universe is a 3-brane embedded in an infinite, non-compactified,
5-dimensional bulk $AdS_5$  spacetime, the $AdS_5$ bulk geometry could then lead
to an exponential suppression of the geometry in distance $w$ away from the
brane and thereby localize gravity to it, thus making it possible for us to
actually be living in a space with an infinite  extra dimension.
Specifically, Randall and Sundrum (RS) considered the case of a static Minkowski
brane at $w=0$ with a cosmological constant $\lambda$ embedded in a
non-compactified $w\rightarrow -w$
$Z_2$ invariant bulk with cosmological constant $\Lambda_5$, and showed that the
5-dimensional  Einstein equations
\begin{equation}
G_{AB}=-\kappa^2_5[\Lambda_5g_{AB}+T_{\mu \nu}\delta^{\mu}_A
\delta^{\nu}_B\delta(w)] 
\label{1}
\end{equation}
(here $A,B=0,1,2,3,5$; $\mu, \nu=0,1,2,3$,
$T^{\mu}_{\phantom{\mu}\nu}=diag(-\lambda,\lambda,\lambda,\lambda)$) had 
solution
\begin{equation}
ds^2(RS)=dw^2+e^{-2\xi|w|/R}(-dt^2+dr^2+r^2d\Omega ) 
\label{2}
\end{equation}
where $2\xi/R=\kappa_5^2
\lambda/3$, $4\xi^2/R^2=-2\kappa^2_5\Lambda_5/3$. A positive sign for $\lambda$ 
then leads to an exponential (warp factor) suppression in distance $w$ away from
either side of the brane of both the geometry$^{\cite{footnote1}}$ and the graviton
propagator \cite{Randall1999a,Randall1999b,Giddings2000},$^{\cite{footnote2}}$ with 
the propagator for widely separated points on the 
brane being found to be \cite{Randall1999a,Randall1999b,Giddings2000} the
standard 4-dimensional $1/q^2$ one. Intriguing as this possibility is, it is 
important to see  just how general it actually is and just how sensitive it is
to the matter  content on the brane, and, indeed, already even in the RS model
itself, no exponential suppression is found in the case where $\lambda$ is
negative  (a growing exponential "antiwarp" case). Consequently, in this paper
we shall explore a different  structure on the brane, one with a more general
perfect fluid brane  energy-momentum tensor
$T^{\mu}_{\phantom{\mu}\nu}=diag(-\rho_b,p_b,p_b,p_b)$  and a geometry [viz.
Robertson-Walker (RW)] lower than the maximally  4-symmetric Minkowski one
considered by Randall and Sundrum, to then find  in the illustrative case of a
brane with non-zero constant spatial  3-curvature $k$ embedded in an
infinite 5-dimensional spacetime endowed with a  cosmological constant, that
there is no exponential suppression of the propagator no matter what the signs
of $\rho_b$ or
$k$.$^{\cite{footnote2a}}$ Exponential suppression is thus seen not to be a generic
property of
$AdS_5$ embeddings.

For the case first of a static, maximally symmetric 3-space embedded
in a general (i.e. not yet $AdS_5$) 5 space,$^{\cite{footnote3}}$ the most general
allowed time independent metric is given by  
\begin{equation}
ds^2=dw^2-dt^2e^2(w)/f(w)+f(w)[dr^2/(1-kr^2)+r^2d\Omega] 
\label{5}
\end{equation}
In such a geometry the Einstein equations take the form 
\begin{eqnarray}
G^{0}_{\phantom{0}0}=-3f^{\prime\prime}/2f+3k/f=\kappa^2_5[\Lambda_5
+\rho_b\delta(w)],
\nonumber \\
G^{1}_{\phantom{1}1}=G^{2}_{\phantom{2}2}=G^{3}_{\phantom{3}3}=
-f^{\prime\prime}/2f-e^{\prime\prime}/e+k/f
=-\kappa^2_5 [-\Lambda_5+p_b\delta(w)],
\nonumber \\
G^{5}_{\phantom{5}5}=-3f^{\prime}e^{\prime}/2fe+3k/f=\kappa^2_5 \Lambda_5,
\label{6}
\end{eqnarray}   
with the most general solution being found to be given by (without loss of 
generality the parameter $\nu$ is taken to be positive)
\begin{eqnarray}
f(w)=\alpha e^{ \nu |w|}+\beta e^{- \nu |w|}-2k/\nu^2,~~
e(w)=\alpha e^{ \nu |w|}-\beta e^{- \nu |w|},
\nonumber \\
e^2(w)/f(w)=f(w)+4k/\nu^2-4(\alpha\beta-k^2/\nu^4)/f(w),~~
\nu=+(-2\kappa^2_5\Lambda_5/3)^{1/2},
\label{7}
\end{eqnarray}   
with the Israel junction 
conditions \cite{Israel1966} at the brane$^{\cite{footnote4}}$ entailing 
\begin{equation}
3\nu(\beta-\alpha)=(\alpha+\beta -2k/\nu^2)\kappa^2_5\rho_b,~~
6\nu(\alpha+\beta)=(\alpha-\beta)\kappa^2_5(\rho_b+3p_b).
\label{8}
\end{equation}   
In general then we see \cite{Mannheim2000} that we get both the warp and
antiwarp factors (rather than just either one of them) in the  
RW case, with 
gravity thus not automatically now being suppressed away from the brane 
despite the presence of a bulk cosmological constant. Moreover, even if we make
the judicious and very specific choice of matter fields
$\kappa^2_5(3p_b+\rho_b)^2+24\Lambda_5=0$, to thereby eliminate the 
$\alpha$ (or the $\beta$)
dependent term in Eq. (\ref{7}) \cite{Mannheim2000}, $f(w)$ will still 
asymptote to a non-vanishing value of 
$-2k/\nu^2$, with non-zero $k$ (of either possible sign) explicitly preventing
localization of gravity.

Now, in reaching this negative result (we show below that the 
result continues to hold even if we take the geometry to be non-static as
well), it is important to note that in the  above solution the bulk geometry is not
necessarily the
$AdS_5$ one that it would have been in the absence of the brane. And, indeed, for a
given bulk geometry to be maximally 5-symmetric, not only must its Einstein tensor 
be proportional to the metric tensor (as is the case away from the brane in Eq.
(\ref{1})), but also its Weyl tensor must vanish identically. However, for the
geometry associated with the general metric of Eq. (\ref{5}) 10 components of
the bulk  Weyl tensor (the 6 $C_{\mu\nu\mu\nu}$ with $\mu\neq\nu$ and the 4
$C_{\mu 5\mu 5}$) are found to be  non-zero, with all of them being found to be
kinematically proportional to
\begin{equation}
C^{12}_{\phantom{12}12}=[-2eff^{\prime \prime}+3ef^{\prime 2}-2efk-
3e^{\prime}ff^{\prime}+2f^2e^{\prime \prime}]/12ef^2.
\label{9}
\end{equation}   
We thus see that in the general solution of Eq. (\ref{7}) the Weyl tensor
is not in fact zero (nor even necessarily asymptotically zero - in the
$\alpha=0$ case for instance, $C^{12}_{\phantom{12}12} \rightarrow
-\kappa^2_5\Lambda_5/6 \neq 0$).$^{\cite{footnote5}}$ Moreover, from the point 
of view of
the matter fields on the brane, there is no apparent reason why the bulk 
geometry should in fact be $AdS_5$ once a lower symmetry (only 
3-symmetric) brane is introduced. However, since it is topologically 
possible to embed the $k>0$ $S_3$ and $k<0$ $R_3$ spaces of constant
spatial 3-curvature into the spatial part of the $R_1\times R_4$ universal
covering  of $AdS_5$, it is possible to find particular values of the metric
coefficients $\alpha$ and
$\beta$ for which the Weyl tensor will in fact vanish, with it in fact being
found to do so in the solution of Eq. (\ref{7}) provided
\begin{equation}
\alpha\beta-k^2/\nu^4=0,
\label{9a}
\end{equation}   
a condition under which the Israel junction conditions then require the
matter fields to obey the very specific constraint  
\begin{equation}
\kappa^2_5\rho_b(2\rho_b+3p_b)=6\Lambda_5,
\label{9b}
\end{equation}   
a constraint which incidentally requires the product $p_b\rho_b$ to
expressly be  negative. Since there is no obvious reason why the bulk and brane
matter fields are obliged to have to be related in this very particular
manner$^{\cite{footnote6}}$  (and even if they were, the
non-vanishing of
$k$ then requires the presence of both the converging and diverging
exponentials), we see that, other than in this very special case, the lowering 
of the symmetry on the brane entails the lowering of the symmetry in the bulk,
and that, regardless of what constraint we may or may not impose on the matter
fields, in no case do we obtain exponential suppression. 

While there would not appear to be any way to avoid this negative result in
general, we note that it might still be possible to do so in the restricted 
case where the post-embedding bulk is in fact taken to be $AdS_5$.
Specifically,  since we would then have two $AdS_5$ patches in  the bulk in
this particular  case, each one of them can then be  brought to the
$ds^2=dw^2+e^{-2\xi w/R}(-dt^2+dr^2+r^2d\Omega )$ form, (i.e. to the form
\begin{equation}
ds^2=dw^2+e^{-2\xi w/R}e^{\pm \nu\eta}(-\nu^2d\eta^2/4+d\chi^2+sinh^2 \chi
d\Omega)
\label{9c}
\end{equation}   
where $r=e^{\pm \nu\eta/2}sinh \chi$, $t=e^{\pm \nu\eta/2}cosh \chi$),
and thus we need to determine which sign of
$\xi$ is then found to ensue for each patch. Thus it could be the case that the
presence of the diverging exponential might only have been due to a
particular choice of coordinates. As we shall see though, this will not in fact
turn out to be the case, though investigation of the issue will prove to be
instructive. 

In analyzing the $\alpha\beta-k^2/\nu^4=0$ case, we note first that since in 
this case the metric coefficients are given by $f(w)=\alpha e^{\nu|w|}+\beta
e^{-\nu|w|} -2k/\nu^2$, $e^2/f=\alpha e^{\nu|w|}+\beta e^{-\nu|w|}+2k/\nu^2$, 
we see that there is no point at which both metric coefficients can
simultaneously vanish. However, it is possible for one to vanish somewhere.
Specifically, if we define 
$\alpha=e^{\sigma-\nu w_0}$, $\beta=e^{\sigma+\nu w_0}$, $e^{2\nu w_0} =
\beta/\alpha $, $e^{\sigma}=|k|/\nu^2$, for $k=+1$ we find that
\begin{equation}
f=(4/\nu^2)sinh^2(\nu w_0/2-\nu|w|/2),~~e^2/f=(4/\nu^2)cosh^2(\nu
w_0/2-\nu|w|/2),
\label{9d}
\end{equation}   
and for $k=-1$ that
\begin{equation}
f=(4/\nu^2)cosh^2(\nu w_0/2-\nu|w|/2),~~e^2/f=(4/\nu^2)sinh^2(\nu
w_0/2-\nu|w|/2).
\label{9e}
\end{equation}   
Thus for $w_0$ negative, all of these metric coefficients will be at their
absolute minimum values at the brane, and thus diverge away from the brane no
matter what the sign of $k$, but for $w_0$ positive, they will all be at  local
maxima at the brane, and actually fall (monotonically) until the points
$w=\pm w_0$ are  reached.$^{\cite{footnote7}}$ Discussion of the various
cases thus depends on whether $w_0$ is positive or negative, i.e. on whether
$\beta/\alpha$ is greater or lesser than one. Since for a Minkowski signatured
metric  the quantity $f(0)=\alpha+\beta -2k/\nu^2$ must necessarily be  positive, we
see from the Israel junction conditions of Eq. (\ref{8}) that $w_0$ is  negative if
$\rho_b$ is negative, while being positive if $\rho_b$ is positive. We must
thus treat these two cases separately, and since for our purposes here it is
sufficient to show that there is at least one case in which there is an
$\xi<0$ antiwarp factor when the RW metric is brought to a form analogous to
that of Eq. (\ref{2}), we shall now investigate this issue in detail in the
negative curvature case.    

While not at all essential in the following, we nonetheless find it convenient 
to normalize the fields so that $f(0)=1$.$^{\cite{footnote8}}$ On restricting now 
to $k=-1$, in the
$\rho_b<0$ case first where the  parameter $w_0$ is given by 
$w_0=-\nu^{-1}L_{+}(1)$ where
\begin{equation}
L_{\pm}(f)=log[\nu^2f/2-1\pm\nu(\nu^2f^2-4f)^{1/2}/2]=-L_{\mp}(f),
\label{11a}
\end{equation}
the coordinate transformation
\begin{eqnarray}
w=[w_0+\nu^{-1}L_{+}(\gamma)]\theta(\gamma-1)-[w_0+\nu^{-1}L_{+}(1/\gamma)]
\theta(1-\gamma),~~r=sinh \chi^{\prime},
\nonumber \\
2t=\{\nu w^{\prime}+log[(\nu^2\gamma-4)/(\nu^2-4)]\}\theta(\gamma-1) +
\{\nu \eta^{\prime}-log[(\nu^2-4\gamma)/(\nu^2-4)]\}\theta(1-\gamma)
\label{11}
\end{eqnarray}   
where $\gamma=e^{\nu\eta^{\prime}-\nu w^{\prime}}$ (and where
$\theta(w_0+\nu^{-1}L_{+}(\gamma))=\theta(\gamma-1)$ for $w_0$ negative)
is found to bring the metric of Eq. (\ref{5}) to the form
\begin{eqnarray}
ds^2=dw^{\prime 2}+[e^{\nu(\eta^{\prime}- w^{\prime})}
\theta(\eta^{\prime}-w^{\prime})+
e^{-\nu(\eta^{\prime}-w^{\prime})}
\theta(w^{\prime}-\eta^{\prime})]
\nonumber \\
\times [-\nu^2d\eta^{\prime 2}/4+d\chi^{\prime 2}+sinh^2 \chi^{\prime}d\Omega].
\label{12}
\end{eqnarray}
Comparing with Eq. (\ref{9c}) 
we thus recognize Eq. (\ref{12}) to precisely possess an antiwarp factor,
just as we had anticipated. Additionally, we also see that in the antiwarp 
factor coordinate system associated with Eq. (\ref{12}) the brane is not at 
rest. Thus once the brane has a matter field source other than a pure 
cosmological constant, it will not be a rest in a pure warp factor or pure 
antiwarp factor coordinate system. Moreover, it is to be expected that the 
brane would not be at rest in Eq. (\ref{12}) since the original $Z_2$ symmetry
was with respect to the coordinate $w$, with there being no $Z_2$ symmetry in
the new coordinate $w^{\prime}$. (Rather the $Z_2$ symmetry of the 
$e^{\nu(\eta^{\prime}- w^{\prime})}
\theta(\eta^{\prime}-w^{\prime})+
e^{-\nu(\eta^{\prime}-w^{\prime})}
\theta(w^{\prime}-\eta^{\prime})$ term is with respect to
the  light cone coordinate $w^{\prime}-\eta^{\prime}$ instead.) It is thus the
absence of any $w^{\prime}\rightarrow -w^{\prime}$ symmetry in the
primed coordinate system which prevents Eq. (\ref{12}) from leading
to localization of gravity.$^{\cite{footnote9}}$      

In the $\rho_b>0$ case where $w_0=\nu^{-1}L_{+}(1)$, 
the coordinate transformation$^{\cite{footnote10}}$  
\begin{eqnarray}
w=[w_0+\nu^{-1}L_{+}(\gamma)]
\theta(w_0+\nu^{-1}L_{+}(\gamma))
+[w_0+\nu^{-1}L_{-}(\gamma)]\theta(w_0+\nu^{-1}L_{-}(\gamma))
\nonumber \\
-[w_0+\nu^{-1}L_{-}(1/\gamma)]\theta(w_0+\nu^{-1}L_{-}(1/\gamma))
-[w_0+\nu^{-1}L_{+}(1/\gamma)]
\theta(w_0+\nu^{-1}L_{+}(1/\gamma)),
\label{13}
\end{eqnarray}   
\begin{eqnarray}
2t=\{\nu
w^{\prime}+log[(\nu^2\gamma-4)/(\nu^2-4)]\}\theta(w_0+\nu^{-1}L_{+}(\gamma)) +
\nonumber \\
\{-\nu
w^{\prime}-log[(\nu^2\gamma-4)/(\nu^2-4)]\}\theta(w_0+\nu^{-1}L_{-}(\gamma))+
\nonumber \\
\{-\nu\eta^{\prime}+log[(\nu^2-4\gamma)/(\nu^2-4)]\}\theta(w_0+
\nu^{-1}L_{-}(1/\gamma))+
\nonumber \\
\{\nu
\eta^{\prime}-log[(\nu^2-4\gamma)/(\nu^2-4)]\}\theta(w_0+
\nu^{-1}L_{+}(1/\gamma))
\label{13a}
\end{eqnarray}   
brings the metric to the form  
\begin{eqnarray}
ds^2=dw^{\prime 2}+\{e^{\nu(\eta^{\prime}- w^{\prime})}\theta(\gamma-4/\nu^2)
[1+\theta(1-\gamma)]+
\nonumber \\
e^{-\nu(\eta^{\prime}-w^{\prime})}
\theta(\nu^2/4-\gamma)[1+\theta(\gamma-1)\}
\times [-\nu^2d\eta^{\prime 2}/4+d\chi^{\prime 2}+sinh^2 \chi^{\prime} d\Omega].
\label{14}
\end{eqnarray}
Thus for $\rho_b$ positive we  again generate an antiwarp
factor, with gravity thus not being localized to the brane for non-zero $k$ 
even for "attractive" $\rho_b$ positive matter sources.$^{\cite{footnote11}}$ 
Consequently, 
having an explicit $AdS_5$ geometry in the bulk is not
sufficient in and  of itself to always guarantee localization of gravity. 
    
\section {Non-static Branes}

For a non-static, maximally symmetric 3-space embedded in a general (i.e. not yet
$AdS_5$) 5 space, the most general allowed metric can be taken to be given by
$^{\cite{footnote12}}$ 
\begin{equation}
ds^2=dw^2-dt^2e^2(w,t)/f(w,t)+f(w,t)[dr^2/(1-kr^2)+r^2d\Omega]. 
\label{19}
\end{equation}
For this metric the components of the Einstein tensor are given by (the dot and the
prime denote derivatives with respect to $t$ and $w$ respectively)  
\begin{eqnarray}
G^{0}_{\phantom{0}0}=-3f^{\prime\prime}/2f+3k/f +3\dot{f}^2/4e^2f,
\nonumber \\
G^{1}_{\phantom{1}1}=G^{2}_{\phantom{2}2}=G^{3}_{\phantom{3}3}=
-f^{\prime\prime}/2f-e^{\prime\prime}/e+k/f
+\ddot{f}/e^2+\dot{f}^2/4e^2f-\dot{e}\dot{f}/e^3,
\nonumber \\
G^{5}_{\phantom{5}5}=-3f^{\prime}e^{\prime}/2fe+3k/f
+3\ddot{f}/2e^2+3\dot{f}^2/4e^2f-3\dot{e}\dot{f}/2e^3,
\nonumber \\
G^{5}_{\phantom{0}0}=3e^{\prime}\dot{f}/2e^3-3\dot{f}^{\prime}/2e^2,
\label{19a}
\end{eqnarray}
while the non-vanishing components of the Weyl tensor (the 6 $C_{\mu\nu\mu\nu}$ 
with $\mu\neq\nu$ and the 4 $C_{\mu 5\mu 5}$) are all found to be
proportional to 
\begin{equation}
C_{0505}=(4e^3ff^{\prime \prime}-6e^3f^{\prime 2}+4e^3fk+
6e^2fe^{\prime}f^{\prime}-4e^2f^2e^{\prime\prime}
-2ef^2\ddot{f}+ef\dot{f}^2+2f^2\dot{e}\dot{f})/8ef^3,
\label{20}
\end{equation}
with the kinematic relation
\begin{equation}
4f^3C_{0505}+2e^2f^2(G^{5}_{\phantom{5}5}-G^{0}_{\phantom{0}0}
-G^{3}_{\phantom{3}3})=6e^2ff^{\prime
\prime}-3e^2f^{\prime 2}
\label{21}
\end{equation}
being found to hold identically (in the coordinate gauge associated with Eq. (19)).

Solving the theory with the same (but now time dependent) sources as in the static 
case discussed earlier is now straightforward. The non-trivial 
($\dot{f}\neq 0$) vanishing of
$G^{5}_{\phantom{0}0}$ entails that
\begin{equation}
e=A(t)\dot{f}
\label{21a}
\end{equation}
where $A(t)$ is an arbitrary function of $t$. Setting 
$G^{0}_{\phantom{0}0}$ equal to $-3\nu^2/2$ in the bulk entails that
\begin{equation}
f^{\prime \prime}=1/2A^2+2k+\nu^2f
\label{22}
\end{equation}
so that
\begin{eqnarray}
f=-1/2A^2\nu^2-2k/\nu^2 +\alpha e^{\nu |w|}+\beta e^{-\nu |w|},
\nonumber \\
e=\dot{A}/A^2\nu^2+A\dot{\alpha}e^{\nu |w|}+A\dot{\beta}e^{-\nu |w|},
\label{23}
\end{eqnarray}
where $\alpha$ and $\beta$ depend on $t$. 
Similarly, setting $G^{5}_{\phantom{5}5}$  equal to $-3\nu^2/2$ in the bulk
entails that 
\begin{equation}
f^{\prime 2}-4kf-\nu^2f^2-f/A^2=B(|w|)
\label{23a}
\end{equation}
where $B(|w|)$ is an arbitrary function of $|w|$ which must be continuous at the
brane.$^{\cite{footnote13}}$ However, compatibility with Eq. (\ref{22}) then entails
that $B$ must actually be constant, with the time dependent functions in the
solution of Eq. (\ref {23}) then being related according to$^{\cite{footnote14}}$  
\begin{equation}
(1+4A^2k)^2/16A^4-\nu^4\alpha\beta=\nu^2B/4. 
\label{24}
\end{equation}
Further, the Israel junction conditions at the brane, yield (see e.g. Binetruy et. 
al. \cite{footnote2a})
\begin{equation}
3\nu(\beta-\alpha)=f(0,t)\kappa^2_5\rho_b,~~
6\nu[\dot{\alpha}-\dot{\beta}]=\dot{f}(0,t)\kappa^2_5(\rho_b+3p_b)
\label{25}
\end{equation}   
(with the standard $\dot{\rho_b}+3\dot{f}(0,t)(\rho_b +p_b)/2f(0,t)=0$ then being
recovered). Thus, just as in the static case, we once again find that in general
(i.e. for completely arbitrary $\rho_b$, $p_b$) there is no asymptotic 
suppression of the metric coefficients.

Moreover, if we additionally require the bulk to be $AdS_5$, the requisite vanishing
of the Weyl tensor then imposes the additional condition  
\begin{equation}
2ff^{\prime \prime}-f^{\prime 2}-\nu^2f^2=0
\label{26}
\end{equation}
in the bulk, 
to require that $B$ actually be zero, with $\alpha$ and $\beta$ then having to be
related to each other according to
\begin{equation}
(1+4A^2k)^2/16A^4=\nu^4\alpha\beta, 
\label{27}
\end{equation}
with the metric coefficient $f(w,t)$ then being given as 
\begin{equation}
f(w,t)=(\alpha^{1/2}e^{\nu |w|/2}- \beta^{1/2}e^{-\nu|w|/2})^2.
\label{28}
\end{equation}
>From Eq. (\ref{27}) we see that product $\alpha\beta$
must thus be non-negative (and even necessarily greater than zero in the $k>0$
case). Hence even in the time dependent case, there is still no exponential
suppression of the geometry for a non-spatially flat brane embedded in an $AdS_5$
bulk.  

The author would like to thank Dr. A. H. Guth for some particularly pertinent
comments and is indebted to Dr. A. Nayeri for his helpful comments. The author 
would also like to thank Drs. R. L. Jaffe and A. H. Guth for the kind 
hospitality of the Center for Theoretical Physics at the Massachusetts 
Institute of Technology where this work was performed. 
This work has been supported in part by funds provided by 
the U.S. Department of Energy (D.O.E.) under cooperative research agreement 
\#DF-FC02-94ER40818 and in part by grant \#DE-FG02-92ER40716.00.


\begin{thebibliography}{99}
\bibitem{footnote1} The essence of the
Randall-Sundrum localization prescription is to locate a brane at the surface
$w=0$ and to piece together two separate  patches of $AdS_5$ spacetime (viz.
patches for each of which the general metric can always be brought to the 
generic form  $ds^2(AdS_5)=dw^2+e^{\pm 2 \xi w/R}(-dt^2+dr^2+r^2d\Omega)$) at
the $w=0$ surface, and to then pick the $AdS_5$ phases in the two
patches in a $Z_2$ invariant way (such as according to $exp(-2\xi
w(\theta(w)-\theta(-w))/R)$) so that there is then a coordinate independent
discontinuity in the brane extrinsic curvature $K_{\mu\nu}(w)$ as the brane is 
crossed from one side to the other. In $Z_2$ invariant geometries it is thus the
coordinate independent sign of this discontinuity which explicitly determines
whether the metric is suppressed or growing in the $w>0^+$ and $w<0^-$ regions, 
with this discontinuity being given by the time-independent and explicitly 
localizing $K_{\mu\nu}(w=0^+)-K_{\mu \nu}(w=0^-)=(-2\xi/R)\eta_{\mu\nu}$ in the
static Minkowski embedded brane case given in Eq. (\ref{2}).
\bibitem{footnote2} With metric
fluctuations of the form $ds^2=dw^2+dx^{\mu}dx^{\nu}(e^{2f(w)}\eta_{\mu
\nu}+h_{\mu\nu})$ obeying $(\partial^2_w-2f^{\prime \prime}-4f^{\prime
2}-e^{-2f}q^{\lambda}q_{\lambda})h_{\mu\nu}(w,q_{\lambda})=0$, localization of
the geometry entails localization of gravity ($h_{\mu \nu}(w,q^2=0)\simeq
e^{2f}$), with the coordinate independent gravitational scattering S-matrix 
then being highly suppressed when it involves a scattering between a particle which
is localized to the brane and one which is not.
\bibitem{footnote3} While the geometry of the bulk might be pure $AdS_5$ 
in the absence of the brane, once the brane is present it is able to set up a
gravitational field of its own in the bulk, to thus potentially lead to a 
bulk geometry other than $AdS_5$ despite the fact that the only matter
field in the bulk itself is a bulk cosmological constant.
\bibitem{footnote4} These conditions take the
generic form $K_{\mu\nu}(w=0^{+})-K_{\mu\nu}(w=0^{-})
=-\kappa^2_5(T_{\mu\nu}-q_{\mu\nu}T^{\alpha}_{\phantom{\alpha}\alpha}/3)$ where 
$q_{\mu \nu}$ is the induced metric on the brane, with the brane extrinsic
curvature having the two independent components 
$K^4_{\phantom {4} 4}=e^{\prime}/e-f^{\prime}/2f$,
$K^1_{\phantom {1} 1}= K^2_{\phantom {2} 2}=K^3_{\phantom {3} 3}=f^{\prime}/2f$
in the RW brane case.
\bibitem{footnote5} This situation  is to be contrasted 
with the  Minkowski case brane considered by Randall and Sundrum themselves,
since the assumption of a Minkowski brane (and thus of a maximally 4-symmetric,
pure cosmological constant perfect fluid on the brane) entails that $e(w)=f(w)$ and
$k=0$, to thus  automatically ensure the kinematic vanishing of the bulk Weyl
tensor (i.e. the metric $ds^2=dw^2+f(w)(-dt^2+dx^2+dy^2+dz^2)$ is kinematically
conformal to flat for any $f(w)$), with the Einstein equations then not only
obliging the bulk geometry to be $AdS_5$ but also forcing one of the two
$\alpha$ and $\beta$ coefficients to vanish, to thereby entail only one 
exponential term.
\bibitem{footnote6} Unless an $AdS_5$ bulk is enforced upon us a priori, say, 
perhaps, by a dimensional compactification of string theory; with the brane
field configuration required for Eq. (\ref{9b}) then possibly being an explicit 
consequence of the embedding, with the brane pressure,
rather than being an a priori input, instead being \cite{Mannheim2000} the output
pressure produced on the brane in response to the pressure generated by 
the gravitational field in the bulk. As regards Eq. (\ref{9b}) we also note in
passing that should it hold, it then has an interesting implication for the
brane cosmology which then ensues. Specifically, it was noted by 
T. Shiromizu, K. Maeda and M. Sasaki,
Phys. Rev. D{\bf 62}, 024012 (2000), that
because of the embedding the leading terms in the Einstein equations on the brane
have to take the form $^{(4)}G_{\mu
\nu}=\kappa^2_5(6\Lambda_5+\kappa^2_5\lambda^2)q_{\mu \nu}/12-8\pi
G_{N}[(\rho_m+p_m)U_{\mu}U_{\nu}+p_mq_{\mu \nu}]$ where 
$G_{N}=\lambda\kappa^4_5/48\pi$ and where we have taken 
the matter density on the brane to be given as $\rho_b=\lambda+\rho_m$, 
$p_b=-\lambda+p_m$. In the presence of Eq. (\ref{9b}) the Einstein tensor on the
brane is then given by
$^{(4)}G_{\mu
\nu}=-8\pi G_{N}[(\rho_m+p_m)U_{\mu}U_{\nu}+p_mq_{\mu \nu} 
-(\rho_m+3p_m)q_{\mu\nu}/2] +O(\kappa^4_5\rho^2)$, with the leading order source
acting just like a perfect fluid with energy density $\rho =
3(\rho_m+p_m)/2$ and pressure $p=-(\rho_m+p_m)/2=-\rho/3$, i.e. just like negative
pressure quintessence.
\bibitem{footnote8} I.e. we pick the fields so
that
$\alpha=[\nu^2+2k \pm \nu(\nu^2+4k)^{1/2}]/2\nu^2$,  
$\beta= [\nu^2+2k \mp\nu(\nu^2+4k)^{1/2}]/2\nu^2$ (i.e. $\nu^2+4k \geq 0)$,  
$\kappa^2_5\rho_b=\mp 3(\nu^2+4k)^{1/2}$, 
$\kappa^2_5p_b=\pm(3\nu^2+8k)/(\nu^2+4k)^{1/2}$, 
$\kappa^4_5\rho_b(\rho_b+p_b)=12k$.
\bibitem{footnote9} With regard to the  coordinate
transformations of Eq. (\ref{11}), we also note in passing that the 
$n^A=n^5\delta^{A}_5$ normal to the brane (as given in the original Eq.
(\ref{5})  coordinate system) transforms so that it acquires components in both
the $w^{\prime}$ and $\eta^{\prime}$ directions, components which then
explicitly become dependent on both $w^{\prime}$ and $\eta^{\prime}$.
Consequently, both the induced  metric
$q_{\mu\nu}dx^{\mu}dx^{\nu}$ on the brane and the brane extrinsic curvature
$K_{\mu\nu}=q^{\sigma}_{\phantom{\sigma}\mu}q^{\tau}_{\phantom{\tau}\nu}
n_{\sigma;\tau}$ acquire $(0,5)$ and
$(5,5)$ components in the transformed coordinate system, to thus enable the
Israel junction  conditions at the brane 
to be compatible with matter field sources more general than the pure brane 
cosmological constant source associated with a brane at rest in a warp factor 
coordinate system. In fact, if the brane were at rest in the Eq. (\ref{2}) 
coordinate system, the junction conditions would then require the brane matter
fluid to be a pure cosmological constant term. Consequently, once the brane 
matter fluid departs from a pure cosmological constant, the brane can not be at 
rest in the warp factor coordinate system, with the brane extrinsic curvature
having to then have an expressly time-dependent discontinuity.
\bibitem{footnote10} For $w_0$ positive the range for
$\theta(w_0+\nu^{-1}L_{+}(\gamma))$ is $\gamma > 4/\nu^2$, and that
of $\theta(w_0+\nu^{-1}L_{-}(\gamma))$ is $1>\gamma >4/\nu^2$.
\bibitem{footnote11} With only the 
flat $k=0$ Robertson-Walker case thus being localizable by an $AdS_5$ embedding
(a case where the vanishing of $C^{12}_{\phantom{12}12}$ entails $e=f$),
it would be of interest to see whether geometries with non-zero $k$ but with a
negligibly small current era  value of $\Omega_k(t)=-kc^2/\dot{R}^2(t)$ could
still admit of an effective  brane-localized current era gravity, with such
localization then having only set in during an early universe inflationary phase.
\bibitem{footnote12} Under the two Hamilton-Jacobi equation type coordinate
transformations  $g^{w w} =g^{\mu
\nu}(\partial w/\partial x_{\mu}) (\partial w/\partial x_{\nu}) =1$, $g^{t w}=
g^{\mu\nu}(\partial t/\partial x_{\mu}) (\partial w/\partial x_{ \nu}) =0$ in the
two dimensional $(x^1,x^2)$ space, the general maximally 3-symmetric
$ds^2=g_{11}(dx^1)^2 +2g_{12}dx^1 dx^2+g_{22} (dx^2)^2 +
f(x^1,x^2)[dr^2/(1-kr^2)+r^2d\Omega]$ metric can be brought to the form exhibited 
in Eq. (\ref{19}). Consequently, the metric coefficient $g_{ww}$ will only depart
from one in fluctuations of the geometry which are less than maximally
3-symmetric.
\bibitem{footnote13} While $B(|w|)$ could in principle take different values on
the two sides of the brane, because it is related to $f^{\prime 2}$ it in fact
does not.
\bibitem{footnote14} Given Eq. (\ref{24}), the one remaining Einstein equation 
(which can conveniently be given as $G^{5}_{\phantom{5}5} - G^{0}_{\phantom{0}0}/2
-3G^{3}_{\phantom{3}3}/2 = 3(ef^{\prime \prime}+fe^{\prime
\prime} - e^{\prime}f^{\prime})/2ef=3\nu^2/2$ in the bulk) is then satisfied
identically by the solution of Eq. (\ref{23}).
\end{thebibliography}
\end{document}